\documentclass[pra,showpacs,floatfix,superscriptaddress,
twocolumn]{revtex4-1}

\usepackage{graphicx}
\usepackage{epstopdf}
\usepackage{amssymb}
\usepackage{amsmath}
\usepackage{dsfont}
\usepackage{xspace}
\usepackage{color}
\usepackage{ulem}
\usepackage{xcolor}
\usepackage{hyperref}
\usepackage{subfigure}

\newcommand{\bra}[1]{\left\langle #1 \right\vert}
\newcommand{\ket}[1]{\left\vert #1 \right\rangle}

\newcommand{\an}{\textcolor{purple}}

\begin{document}

\title{Kernel-based quantum regressor models learn non-Markovianity}
\author{Diego Tancara}
\affiliation{Centro de \'Optica e Informaci\'on Cu\'antica, Universidad Mayor, Santiago, Chile}
\author{Hossein T. Dinani}
\affiliation{Escuela Data Science, Facultad de Ciencias, Ingenería y Tecnología, Universidad Mayor, Santiago, Chile}
\author{Ariel Norambuena}
\affiliation{Centro de \'Optica e Informaci\'on Cu\'antica, Universidad Mayor, Santiago, Chile}
\author{Felipe F. Fanchini}
\affiliation{Faculdade de Ci\^encias, UNESP - Universidade Estadual Paulista, Bauru, SP, 17033-360, Brazil}
\author{Ra\'ul Coto}
\email{raul.coto@protonmail.com}
\affiliation{Department of Physics, Florida International University, Miami, Florida 33199, USA}
\date{\today}

\begin{abstract}
Quantum machine learning is a growing research field that aims to perform machine learning tasks assisted by a quantum computer. Kernel-based quantum machine learning models are paradigmatic examples where the kernel involves quantum states, and the Gram matrix is calculated from the overlap between these states. With the kernel at hand, a regular machine learning model is used for the learning process. In this paper we investigate the quantum support vector machine and quantum kernel ridge models to predict the degree of non-Markovianity of a quantum system. We perform digital quantum simulation of amplitude damping and phase damping channels to create our quantum dataset. We elaborate on different kernel functions to map the data and kernel circuits to compute the overlap between quantum states. We show that our models deliver  accurate predictions that are comparable with the fully classical models.    

\end{abstract}

\maketitle

\section{Introduction.}

During the last decades we have witnessed the rapidly growing fields of Artificial Intelligence (AI) and Quantum Computing (QC). The basis for AI and QC were developed in the past century. However, it is now that this knowledge is widely available for research, business, health, among others. AI aims to provide machines with human-like intelligence. From the very beginning, AI has been conceived in different ways, leading to the development of different branches, known as Machine Learning (ML)\cite{scikit-learn,Mehta2019,Carleo2019}, Deep Learning~\cite{Paszke2017} and Reinforcement Learning~\cite{Sutton1998}. ML is based on statistical learning, where the machine learns from data that has already been labelled (Supervised learning) or from unlabelled data (Unsupervised learning). In recent years, Supervised learning has undoubtedly impacted on physics~\cite{Mehta2019,Carleo2019,Dunjko2018}. In particular, it is known for unravelling patterns from datasets that yield quantum phase transitions~\cite{Carrasquilla2017,Canabarro2019aa}. 
         
Quantum computing is also at the forefront of current technologies. Nowadays, research groups have delivered highly functional and fault-tolerant quantum algorithms encompassing a wide variety of systems including: superconducting qubits~\cite{Kandala2017,Arute2019}, trapped ions\cite{Wright2019},  cold atoms~\cite{Graham2022}, photonics~\cite{Peruzzo2014,Arrazola2019} and color centers in diamond~\cite{Abobeih2022}. In the last years, quantum computers have pushed further the boundaries of physics, chemistry, biology, and computing itself, with groundbreaking achievements in the simulation of novel materials~\cite{Babbush2018}, molecules~\cite{Peruzzo2014,OMalley2016,Kandala2017,Nakanishi2019}, in designing algorithms towards quantum supremacy~\cite{Peng2008,Arute2019} and quantum machine learning~\cite{Rebentrost2014,Wiebe2015,Cai2015,Li2015,Biamonte2017,Havlicek2019,Schuld2019,He2019,Mengoni2019,Bartkiewicz2020,Johri2021,Willsch2020,Zhang2020,Park2020,Khan2020,Schuld2021,Goto2021,Wang2021,Gyurik2021,Saeedi2021,Ding2021}.

Among the main obstacles to be overcome in the development of quantum technologies is the interaction of the quantum system with the environment. This interaction disturbs the quantum state and, in general, can be divided into two types of processes: Markovian and non-Markovian \cite{BreuerPet}. Non-Markovian processes are those in which memory effects are taken into account and their importance can be noted in several processes and protocols such as state teleportation \cite{laine2014}, quantum metrology \cite{chin2012} and even in current quantum computers \cite{white2020}. In this paper we use quantum machine learning to determine the degree of non-Markovianity of a quantum process. We focus on kernel-based machine learning models to learn from quantum states. Our results shows that the quantum computer can create the dataset, but also treat and learn from it, providing feedback on the very process in which it is involved. 

The paper is organized as follows. In Sec.~\ref{Sec_QSVM} we introduce two quantum machine learning models based on kernels, namely: Quantum Support Vector Machine and Quantum Kernel Ridge models. The goal of these models is to estimate the degree of non-Markovianity from a dataset made of quantum states. Furthermore, we elaborate on the performance of the models based on three different kernel functions and four different kernel circuits to measure the overlap between two quantum states. All these possible combinations yield different Gram matrices. In Sec.~\ref{Sec_DQS}, we introduce the Digital Quantum Simulation approach that we followed to describe the evolution of the system in Amplitude Damping and Phase Damping channels. In Sec.~\ref{Sec_results}, we show our main results regarding the prediction of the degree of non-Markovianity. In Sec.~\ref{Sec_conclusions} we deliver the final remarks of this work.


\section{Kernel-based machine learning models.}\label{Sec_QSVM}

Quantum machine learning aims to perform machine learning tasks assisted by a quantum computer. In recent years, different implementations have been addressed, including Variational Quantum Circuits~\cite{Benedetti2019,Cerezo2021,Dinani2022}, quantum Nearest-Neighbor methods~\cite{Wiebe2015} and quantum Kernel Methods~\cite{Rebentrost2014,Li2015,Schuld2021}. The latter, naturally appears in models that support a kernel function to represent the data into a feature space. Two well-understood examples are the Support Vector Machine (SVM) and the Kernel Ridge Regressor (KRR) models. Their extension to the quantum domain via a precomputed kernel is straightforward. Next, we describe the SVM and KRR models and their connection with the kernel. 

\subsection{Support Vector Machine}

One of the most broadly used models in ML is Support Vector Machines (SVM)~\cite{Vapnik1995}. This model can be used for classification~\cite{Burges1998,Opper2001} and regression~\cite{Vapnik1995,Scholkpf2000,Smola2004} tasks. The former, gives rise to an intuitive representation that relies on a hyperplane that splits the dataset into different classes. Therefore, predicting the label of unknown data only depends on where the data samples fall regarding the hyperplane. In general, other models also use a hyperplane. However, the SVM sets the maximum-margin, i.e. maximizing the distance between the hyperplane and some of the boundary training data, which are the data samples close to the edge of the class. These particular samples are known as support vectors (SVs). Since SVs are a subset of the training dataset, this model is suitable for situations where the number of training data samples is small compared to the feature vector's dimension. Once the model has fitted the training dataset, it can be used as a decision function that predicts new samples, without holding the training dataset (eager learning algorithm) in memory. In this work we will focus on a regression task, which predicts a real number rather than a class. In what follows, we briefly describe the mathematical formulation of the optimization problem. More details can be found in Ref.~\cite{Fanchini2021}.

SVM delivers the tools for finding a function $f(\vec{x})$  that fits the training dataset $\{\vec{x}_i,y_i \}$, where $\vec{x}_i\in \mathbb{R}^d$ are the feature vectors with dimension $d$, and $y_i\in\mathbb{R}$ are the corresponding labels. Note that $i$ runs over the number of training samples ($i=1,2,\dots,\mathit{l}$). We begin with the linear function $f(\vec{x})=\vec{w}\cdot\vec{x} +b$, with $\vec{w}\in \mathbb{R}^d$ and $b\in \mathbb{R}$ being fitting parameters. We shall discuss the case of nonlinear separable data later on. For $\epsilon$-SVM~\cite{Vapnik1995}, deviations of $f(\vec{x})$ from the labeled data ($y_i$) must be smaller than $\epsilon$, i.e. $\vert f(\vec{x})-y_i \vert\leq\epsilon$. Moreover, we must address the model complexity as given by the $l_2$-norm $\left\Vert\vec{w}\right\Vert^2$, and the tolerance for deviations $\xi_i,\xi_i^{\ast}$ (slack variables) larger than $\epsilon$, that are weighted by $C>0$. Therefore, the optimization problem can be stated as~\cite{scikit-learn,Vapnik1995,Smola2004},
\begin{eqnarray}
&\mbox{minimize} \hspace*{1cm}  \frac{1}{2} \left\Vert \vec{w}\right\Vert ^2 + C\sum_i \left( \xi_i + \xi_i^\ast \right) \nonumber\\
&\mbox{subjected to} \hspace*{1cm} \left\lbrace \begin{array}{l}
  y_i - \vec{w}\cdot\vec{x}_i -b \leq \epsilon + \xi_i \\
  \vec{w}\cdot\vec{x}_i +b - y_i\leq \epsilon + \xi_i^\ast \\
 \xi_i,\xi_i^\ast \geq 0
\end{array} \right. 
\end{eqnarray}

One can solve this problem introducing the Lagrange multipliers $\alpha_i,\alpha^\ast_i,\eta_i,\eta^\ast_i \geq 0$, with the Lagrangian defined as~\cite{Vapnik1995,Smola2004,Scholkpf2000},
\begin{align}\label{lagrangian}
L &=\frac{1}{2}\left\Vert \vec{w}\right\Vert^2 + C\sum_i \left( \xi_i + \xi_i^\ast \right) - \sum_i (\eta_i\xi_i +\eta_i^\ast\xi_i^\ast)\nonumber \\
& -\sum_i\alpha_i(\epsilon + \xi_i -y_i + \vec{w}\cdot\vec{x}_i +b)\nonumber \\
& -\sum_i\alpha_i^\ast (\epsilon + \xi_i^\ast + y_i -\vec{w}\cdot\vec{x}_i -b).
\end{align}

From the vanishing partial derivatives $\partial_bL$, $\partial_wL$, $\partial_{\xi}L$ and $\partial_{\xi^\ast}L$ the optimization problem can be recast as,
\begin{align}
&\mbox{maximize} \hspace*{1cm}  \left\lbrace \begin{array}{l}-\frac{1}{2}\sum_{i,j} (\alpha_i-\alpha_i^\ast)(\alpha_j-\alpha_j^\ast)\langle x_i,x_j\rangle \\
-\epsilon\sum_i (\alpha_i + \alpha_i^\ast) + \sum_i y_i(\alpha_i - \alpha_i^\ast) \end{array}\right. \nonumber  \\
&\mbox{subjected to} \hspace*{1cm} \left\lbrace \begin{array}{l}
   \sum_i (\alpha_i-\alpha_i^\ast)=0 \\
  \alpha_i,\alpha_i^\ast\in[0,C]\end{array}\right. 
\end{align}

For convenience, we have written the dot product as an inner product, $\langle x_i,x_j\rangle=\vec{x}_i\cdot\vec{x}_j$. From $\partial_wL=0$ we find $\vec{w}=\sum_i(\alpha_i - \alpha_i^\ast)\vec{x}_i$, that leads to the decision function 
\begin{equation}\label{f_decision}
f(\vec{x})=\sum_i(\alpha_i - \alpha_i^\ast)\langle x_i,x\rangle + b,
\end{equation}

that depends on the inner product between the unlabeled data ($\vec{x}$) and the training data ($\vec{x}_i$). We can recover $b$ from the Karush-Kuhn-Tucker (KKT) condition, which states that at the solution point of the Lagrangian, the product between the Lagrange multipliers and the conditions vanishes. We remark that this calculation is computed internally in \textit{scikit-learn} library~\cite{scikit-learn}. We would like to stress that the decision function in Eq.~\eqref{f_decision} has a sparse representation in terms of $\alpha_i,\alpha_i^\ast$. Only a small subset of the training dataset (support vectors) contributes to the decision function. In Appendix~\ref{Appendix_svm}\an{,} we show the arguments for the sparsity and the calculation of $b$.

We have introduced so far a linear decision function that can handle linearly separated data. For nonlinearly separated data, it is possible to define a clever kernel function $k(x_i,x)$ that generalizes $\langle x_i,x\rangle$ by taking the samples to a higher dimensional space, where they are linearly separable. We elaborate further on this idea later on. 

\subsection{Kernel Ridge Regressor}

Kernel Ridge Regressor (KRR) is another important nonlinear machine learning model. It has been successfully used to predict the evolution of quantum systems~\cite{Rodriguez2022}. It combines Ridge Regression with the kernel trick \cite{scikit-learn,Hastie2009}. The former, provides a linear solution based on least squares with $l_2$ regularization that penalizes large coefficients. Like in SVM, the $l_2$-norm prevents model complexity, while the kernel allows the model to learn a nonlinear function in the original space. This model offers a straightforward optimization problem stated by \cite{scikit-learn} 
\begin{equation}\label{rr}
\mbox{minimize} \hspace*{1cm}  \sum_{i=1}^N \left\Vert \vec{w}\cdot\vec{x}_i-y_i\right\Vert ^2 + \alpha  \left\Vert \vec{w} \right\Vert. 
\end{equation}

The above problem can be written in an equivalent way as~\cite{Hastie2009},
\begin{align}\label{rr_dual}
&\mbox{minimize} \hspace*{1cm}  \sum_{i=1}^N \left( y_i - \vec{w}\cdot\vec{x}_i - b \right)^2, \nonumber\\
&\mbox{subjected to} \hspace*{1cm} \left\Vert \vec{w} \right\Vert^2\leq \alpha_d,
\end{align} 

where there is a one-to-one correspondence between the hyperparameters $\alpha$ and $\alpha_d$. Introducing the Lagrange multipliers as in the previous subsection the decision function can be found as,
\begin{equation}\label{f_decision_krr}
f(\vec{x})=\sum_i\beta_i k(x_i,x) + b,
\end{equation}

It is worth noting that SVM and KRR are similar in terms of the $l_2$ regularization and that both use the kernel trick, but the loss function is different. While SVM relies on a linear $\epsilon$-insensitive loss, KRR uses squared error loss. The former implies that all the training points that result in errors that fall inside the $\epsilon$-tube do not contribute in the solution, which originates sparseness. In contrast, KRR considers all the training points. This yields differences in the performance of these models.     

Machine learning algorithms have greatly profited from kernel functions~\cite{Dunjko2018,Mengoni2019,Schuld2021}. Therefore, we now introduce a generalization of the decision function to learn from nonlinear data. The kernel can be understood as a measure of similarities between two vectors, and it supports representations ranging from polynomial to exponential functions~\cite{scikit-learn}. Along this paper we consider three different functions for the kernel $k(x_i,x_j)$, namely: linear $\langle x_i,x_j\rangle + c$, polynomial $\left(\langle x_i,x_j\rangle +c\right)^d$, and exponential $\exp(-\sigma\sqrt{1-\langle x_i,x_j\rangle})$. 

We have so far addressed the classical part (optimization problem) of this hybrid quantum machine learning approach. In the next subsection we will focus on implementing the kernel through a quantum circuit.

\subsection{Quantum Kernels}

We have noted that the kernel provides efficient separability in nonlinear regions. The main idea behind the kernel is that it allows to map the data to a higher-dimensional space, termed as “featured space”~\cite{Fanchini2021}. In general lines, let’s consider a feature map  $\phi:x\in\chi\rightarrow\phi(x)\in H$ that encodes information from a certain domain $\chi$ (commonly $\chi\in R^n$) to a feature space $H$. The advantages of using the map rely on the “kernel trick”~\cite{Dunjko2018}, which allows us to set the decision function without the explicit calculation of $\phi(x)$. This idea has encouraged researchers to bridge classical and quantum machine learning \cite{Havlicek2019,Schuld2019,Schuld2021}. Let’s consider a Hilbert space $H$ that contains the states of a quantum system. Now, instead of encoding the information of $\chi$ in a feature space given by functions $\phi(x)$, with $x\in\chi$, the information is encoded in quantum states $\ket{\phi(x)}\in H$~\cite{Schuld2021,LaRose2020,Weigold2021}, which is known as quantum embedding.  Quantum embedding is a crucial step in the process and, in some cases, may lead to a disadvantage against classical models. To overcome this, we resorted to perform digital quantum simulation of the quantum dynamics rather than classical simulation~\cite{Fanchini2021}, which allows us to handle quantum states to build up the kernel. Thus, we train our model with a symmetric and semi-positive definite matrix (Gram matrix), rather than the data samples (quantum states). 

The next step is to calculate the kernel from the training samples $\rho_i$. A natural choice is the pairwise trace distance between the quantum states ($\mbox{Tr}[\rho_i\rho_j]$), that is commonly carried by the Swap test~\cite{Smolin1996,Cincio2018}. In what follows we describe the circuit implementation. First, we encode the information into two different qubits. Each of these qubits undergoes a NM evolution (induced by independent ancilla qubits). Then the overlap between states $\rho_i$ and $\rho_j$ yields the matrix element $k(\theta_i,\theta_j) = \mbox{Tr}[\rho_i\rho_j]$, where $\theta_i$ is the parameter that control the NM evolution. We note that for the case of pure states, $\rho_i=\vert\psi_i\rangle\langle\psi_i\vert$ and $\rho_j=\vert\psi_j\rangle\langle\psi_j\vert$, the kernel simply reduces to $\vert\langle\psi_i\vert\psi_j\rangle\vert^2$. 

We describe next different implementations for the overlapping.

\subsubsection{Swap test}

 The Swap test is a high level sequence of quantum operations that involves two data qubits, an ancilla qubit, two-qubits (CNOT) gates, one-qubit gates and a final measurement on the ancilla~\cite{Smolin1996}, see Fig.~\ref{fig:kernel}. By measuring the probability of finding the ancilla in state $\ket{0}$ ($P_0$), one obtains the state overlapping by computing $\mbox{Tr}[\rho_i\rho_j] = 2P_0-1$.

\subsubsection{Inversion test}

Our second kernel considers the quantum state of a closed system (unitary evolution), that encompasses the system qubit and the environment ancilla qubit~\cite{Lloyd2020}. It begins with two different quantum states driven by an unitary evolution $U(\theta)$, such that $\ket{\Psi_{\theta}}=U(\theta)\vert00\rangle$, with $\vert00\rangle = \ket{0}_s\otimes\ket{0}_a$. The kernel is defined as the squared absolute value of the projection between these two states, that is equivalent to two subsequent evolutions--- assuming that the inverse evolution $U^{\dag}(\theta_i)$ can be implemented. The matrix elements reads, 
\begin{eqnarray}
	k(\theta_i,\theta_j) &=& |\bra{\Psi_{\theta_i}}\Psi_{\theta_j}\rangle|^2 = |\bra{00}U^{\dag}(\theta_i)U(\theta_j)\ket{00}|^2\nonumber \\
		&=& |\bra{00}\Theta\rangle|^2,
\end{eqnarray}

\begin{figure*}[ht]
	\centering
	\includegraphics[width=0.7 \textwidth]{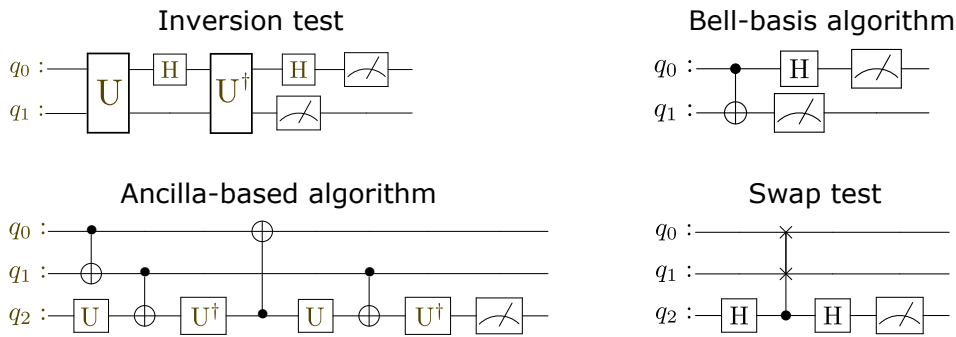}
	\caption{ Quantum circuits compute the overlap between two quantum states in the kernel function to calculate the Gram matrix. For the inversion test $U$ represents either the amplitude damping or phase damping channel depicted in Fig.~\ref{fig:channels_circ}. For the ancilla-based algorithm (ABA) $U=T^\dagger H$~\cite{Cincio2018}.}
	\label{fig:kernel}
\end{figure*}

where $\ket{\Theta} = U^{\dag}(\theta_i)U(\theta_j)\ket{00}$. In contrast to the Swap test kernel, this one requires two measurements, which allows us to decrease the number of quantum registers (Fig.~\ref{fig:kernel}). We remark that this kernel is not experimentally feasible for the particular goal of detecting non-Markovianity. In general, one has no access to perform measurements upon the environment. In addition, it requieres reverse unitary interactions of the system-environment dynamics. Nevertheless, we consider it because it may be applied to other machine learning tasks~\cite{Lloyd2020} and it delivers the best accuracy we found in this paper.

\subsubsection{Ancilla-based algorithm}

The Ancilla-based algorithm (ABA) is a variation of the Swap test that conveniently reduces the number of gates. It was first discovered in the context of quantum optics~\cite{Garcia2013}, and rediscovered later with assistance of a neural network and introduced for quantum circuits~\cite{Cincio2018}. The circuit is depicted in Fig.~\ref{fig:kernel}.

\subsubsection{Bell-basis algorithm}

The Bell-basis algorithm (BBA) considers less resources than the previous one (ABA), but demands Bell-basis measurements on all the system qubits~\cite{Cincio2018}. The circuit is depicted in Fig.~\ref{fig:kernel}. 

In this paper we do not intent to explicitly compare the accuracy of all these approaches for estimating the overlapping (for a comparison between Swap test, ABA and BBA see~\cite{Cincio2018}). We will compare them in terms of the accuracy of the decision function. 

In the next section we describe the quantum circuits that account for the interaction between the system qubit with the environment ancilla that ultimately yields non-Markovianity.

\section{Digital Quantum Simulation of non-Markovian channels}\label{Sec_DQS}
The main purpose of this paper is to determine the degree of non-Markovianiaty of a quantum process using a quantum machine learning algorithm. We begin simulating two non-Markovian channels, amplitude damping and phase damping, whose degree of non-Markovianity can be controlled. For this purpose we simulate the processes using usual circuit routines, taking auxiliary qubits to represent the environment. In this section, we show how the degree of non-Markoviniaty is calculated and present how the non-Markovian amplitude damping and phase damping processes can be simulated using a quantum circuit.

\subsection{Calculating the degree of non-Markovianity}

There are different ways to measure the degree of non-Markovianity. The most popular measures are based on the trace distance dynamics~\cite{Breuer2009}, the dynamics of entanglement~\cite{Rivas2010,Chruscinski2011}, and mutual information \cite{Luo2012}, among others~\cite{Pollock2018}. In this paper we consider the measure based on entanglement dynamics of a bipartite quantum state that encompasses the system that interacts with the environment and an ancilla qubit that is isolated from it~\cite{Rivas2010}. Worthwhile noticing that this ancilla only serves the purpose of quantifying non-Markovianity and it is not implemented in the quantum circuits, in contrast to the ancilla used to simulate the effect of the environment for the amplitude damping and phase damping processes. 

A monotonic decrease in the entanglement of the bipartite system implies that the dynamics is Markovian. An increase in the entanglement during the evolution is a result of memory effects and thus non-Markovianity. The measure can be calculated as 
\begin{equation}
    \mathcal{N}={\text {max}}\int_{dE(t)/dt>0}{\frac{dE(t)}{dt}dt},
\end{equation}
where the maximization is done over all initial states and $E$ is the measure of entanglement. 
It has been found that the maximization is achieved for Bell states \cite{Neto2016}. Therefore, we consider a bipartite system in a Bell state and use concurrence as the measure of entanglement \cite{Hill1997}.

\subsection{Amplitude Damping}\label{sub_AD}

For the amplitude damping (AD) channel, we consider a qubit interacting with a bath of harmonic oscillators, given by the Hamiltonian ($\hbar=1$) \cite{Hakkika, Whalen}
\begin{eqnarray}
&&H=\omega_0 \sigma_{+} \sigma_{-}+\sum_{k}{\omega_k a^{\dagger}_k a_k} \nonumber\\
&&\qquad+\sum_k (g^{*}_k \sigma_{+}a_k+g_k \sigma_{-}a^{\dagger}_k).
\end{eqnarray}
Here, $\sigma_{+}=\sigma^{\dagger}_{-}=|1 \rangle\langle 0|$ with $|1\rangle$ ($|0\rangle$) corresponding to the excited (ground) state of the qubit with transition frequency $\omega_0$, $a_k (a^{\dagger}_k)$ is the annihilation (creation) operator of the $k$-th mode of the bath with frequency $\omega_k$, and $g_k$ is the coupling between the qubit and the $k$-th mode. We assume that the bath has a Lorentzian spectral density 
\begin{equation}
    J(\omega)=\frac{1}{2\pi}\frac{\gamma_0 \lambda^2}{(\omega_0-\omega)^2+\lambda^2},
\end{equation}
where $\lambda \approx 1/\tau_r$ with $\tau_r$ being the environment correlation time, $\gamma_0\approx 1/\tau_s$ where $\tau_s$ is the typical time scale of the system.

The dynamics of the qubit that is coupled resonantly with the environment can be expressed as 
\begin{equation}
    \rho(t)=\sum_{i=0}^{1}{M_i(t)\rho(0)M^{\dagger}_i(t)},
\end{equation}
where the Kraus operators are given by \cite{Nielsen} \cite{Garcia-Perez}
\begin{eqnarray}
    &&M_0 (t)=|0\rangle\langle0|+\sqrt{p(t)}|1\rangle\langle 1|, \\
    &&M_1 (t)=\sqrt{1-p(t)}|0\rangle \langle 1|,
\end{eqnarray}
in which
\begin{equation}\label{eq:ptAD}
    p(t)=e^{-\lambda t}\left[\frac{\lambda}{d}\sinh(d t/2)+\cosh(d t/2)\right]^2,
\end{equation}
with $d=\sqrt{\lambda^2-2\gamma_0\lambda}$. The dynamics is known to be non-Markovian in the strong coupling regime $\lambda<2\gamma_0$ $(\tau_s<2\tau_r)$ \cite{Bellomo}. 

\begin{figure}[h]
\centering
    \includegraphics[width=0.3\textwidth]{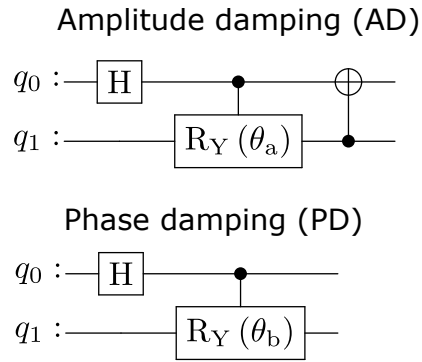}
   \caption{Quantum circuits for simulating AD and PD channels.}
   \label{fig:channels_circ}
\end{figure}

The AD process can be simulated for a general scenario with a quantum circuit via an ancilla qubit \cite{Nielsen,Garcia-Perez}. After tracing out the ancilla qubit we obtain the desired mixed state. Figure \ref{fig:channels_circ} shows the quantum circuit. The Hadamard gate prepares the qubit in the superposition state $\left(|0\rangle+|1\rangle\right)/\sqrt{2}$ while the controlled rotation and CNOT gates simulate the interaction of the qubit with the environment. In this circuit, the angle $\theta_a$ is given by \cite{Nielsen, Garcia-Perez}
\begin{equation}\label{eq:thetaad}
    \theta_a=2\arccos\left(\sqrt{p(t)}\right),
\end{equation}
where $p(t)$ is given in Eq.~\eqref{eq:ptAD}.

\subsection{Phase Damping}

For the phase damping (PD) channel, following Ref.~\cite{daffer04}, we consider a qubit undergoing decoherence induced by a colored noise given by the time dependent Hamiltonian ($\hbar=1$)
\begin{equation}
H(t)=\Gamma(t) \sigma_z.
\end{equation}
Here, $\Gamma(t)$ is a random variable which obeys the statistics of a random telegraph signal defined as $\Gamma(t)=\alpha (-1)^{n(t)}$, where $\alpha$ is the coupling between the qubit and the external influences, $n(t)$ is a random variable with Poisson distribution with mean $t/(2\tau)$, and $\sigma_z$ is the Pauli $z$ operator. In this case, the dynamics of the qubit is given by the following Kraus operators \cite{daffer04}
\begin{eqnarray}
    &&M_0(t)=\sqrt{\frac{1+\Lambda(t)}{2}}\mathbb{I}, \\ &&M_1(t)=\sqrt{\frac{1-\Lambda(t)}{2}}\sigma_z,
\end{eqnarray}
where
\begin{equation}\label{eq:LambdaPD}
\Lambda(t)=e^{-t/(2\tau)}\left[\cos(\frac{\mu t}{2\tau})+\frac{1}{\mu}\sin(\frac{\mu t}{ 2\tau})\right],
\end{equation}
with $\mu=\sqrt{(4 \alpha \tau)^2-1}$, and $\mathbb{I}$ being the identity matrix. 

For $\alpha \tau>1/4$ the dynamics is non-Markovian, while for $\alpha \tau<1/4$ it is Markovian. The PD channel can be simulated using a quantum circuit, shown in Fig.~\ref{fig:channels_circ}~\cite{Nielsen}. In this circuit, the Hadamard gate prepares the qubit into the superposition state and the controlled rotation simulates the interaction with the environment. The angle $\theta_p$ is given by
\begin{equation}\label{eq:thetapd}
    \theta_p=2\arccos\left(\Lambda(t)\right),
\end{equation}
where $\Lambda(t)$ is given in Eq.~\eqref{eq:LambdaPD}.

\section{Results}\label{Sec_results}

\begin{figure*}
	\centering
	\includegraphics[width=0.7 \textwidth]{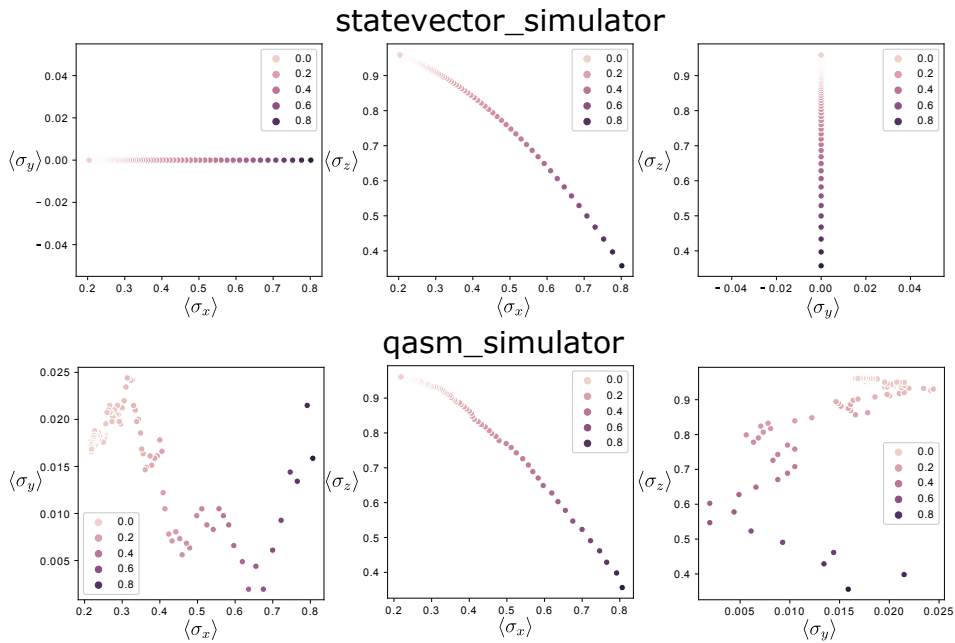}
	\caption{Expectations values delivered by the noisy \textit{qasm\_simulator} exhibit small dispersion given by the shot-noise, in contrast to the ideal statevector\_simulator. We only observe correlations in the plane defined by $\sigma_x$ and $\sigma_z$.}
	\label{fig:exp_val}
\end{figure*}

We perform our simulations with the \textit{statevector\_simulator} and \textit{qasm\_simulator}, integrated in the Aer's package from IBM qiskit~\cite{qiskit}. For comparison, we also run simulations using Pennylane library~\cite{Bergholm2018}, obtaining similar outcomes. The \textit{statevector\_simulator} is an ideal simulator that considers the evolution of the wavefunction. In contrast, the \textit{qasm\_simulator} mimics the open dynamics of the IBM quantum computer. This means that it considers losses and shot-noise. However, it allows us to set all qubits equal and fully connected (not relying on a specific quantum hardware).

It is well-known that the quantum state of a qubit can be represented as a point in a sphere of radius one (Bloch's sphere). A generic state can be represented in the Bloch's sphere in terms of the expectations values as 
\begin{equation}
\rho=\frac{1}{2}\left(\mathbb{I} + \sum_{i=x,y,z}\langle\sigma_i\rangle\sigma_i\right),    
\end{equation}

where $\mathbb{I}$ is the $2\times2$ identity matrix. 

For illustration we firstly focus on the amplitude damping channel. In Fig.~\ref{fig:exp_val} we show the expectation values calculated using the \textit{statevector\_simulator} and \textit{qasm\_simulator}. The former, provides outcomes with no dispersion (top), as expected from the ideal simulation. On the other hand, \textit{qasm\_simulator} delivers more realistic results that include dispersion (bottom). This dispersion will be pivotal for selecting the best algorithm that computes the overlap, since \textit{statevector\_simulator} brings no significant difference in the prediction. In order words, simulations on \textit{statevector\_simulator} may be misleading when selecting a machine learning model.

\begin{figure}
	\centering
	\includegraphics[width=0.45 \textwidth]{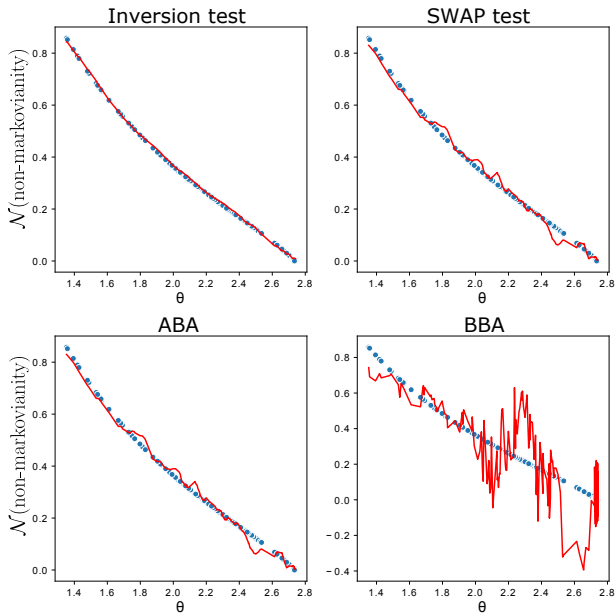}
	\caption{QSVM prediction of non-Markovianity as a function of the rotation angle $\theta$ for different kernel circuits. The inversion test outperforms the others. We set the hyperparameters $\{C=0.5,\epsilon=0.01 \}$.}
	\label{fig:kernels}
\end{figure}

In Fig.~\ref{fig:kernels} we show the degree of NM for the amplitude damping channel as a function of the parameter $\theta$ (rotation angle that controls NM introduced in subsection~\ref{sub_AD}). For the calculations, we used \textit{qasm\_simulator} with the exponential kernel function ---that yields the best accuracy as shown in Appendix~\ref{Appendix_kernel_func}. For exploration of the algorithms we only focus on QSVM. We manually seek optimal hyperparameters and report the prediction on the training dataset. A more robust analysis will be given later on. We can observe that the inversion test leads to a feature space that allows better prediction of the degree of NM. 

\begin{figure}
	\centering
	\includegraphics[width=0.40 \textwidth]{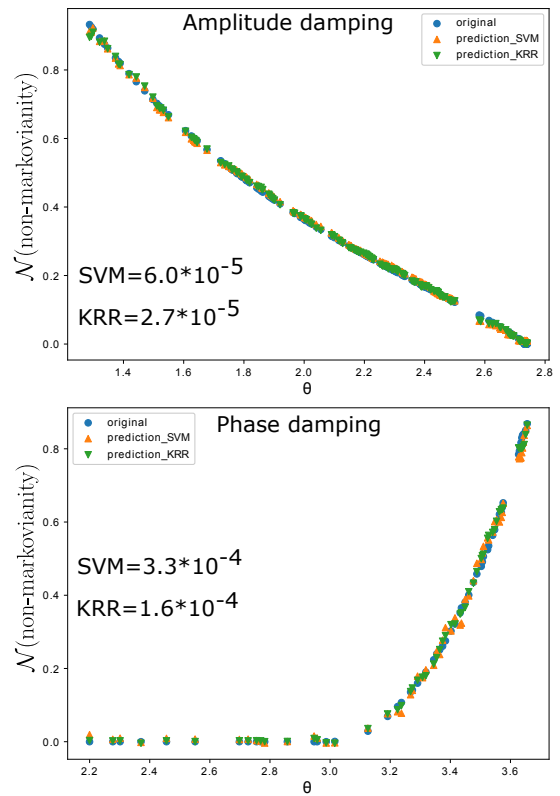}
	\caption{Both QSVM and QKRR deliver accurate predictions of the degree of non-markovianity, based on the mean squared error score. For a small training dataset QSVM performs better (not shown here). For a sufficiently large number of points QKRR provides a smaller mean squared error.   }
	\label{fig:krr}
\end{figure}

We now compare the performance between QSVM and QKRR. Hereafter, we focus on simulations on the \textit{qasm\_simulator} for the inversion test with exponential function. To prevent overfitting, we use two steps for cross-validation. First, we use the \textit{train\_test\_split} function in scikit-learn~\cite{scikit-learn} to randomly split the training set from the test set. Then, we use the \textit{GridSearchCV} function to explore the best fitting hyperparameters for each model, and we use a five-fold cross-validation. Thus, \textit{GridSearchCV} provides the best estimator for the range of given parameters averaged over five different sampling of the training set. Finally, we used these estimators to predict the test set, which contains the data that the model has not seen. In Fig.~\ref{fig:krr} we show our predictions for amplitude damping and phase damping. One can observe that both models succeeded in predicting the degree of non-Markovianity, besides small differences in the score (mean squared error). However, there are important aspects that might be taken into account before selecting one over the other. First, we remark that QSVM requires less training data to deliver good fittings. This is known, and it results from the sparseness in the training samples (only SVs contribute). Therefore, QSVM provides a major advantage given that the most time consuming operation is the calculation of the Gram matrix. Thus, less training samples reduces the overall computation time. In contrast, we observe that as the number of data samples increases, QKRR improves.

For comparison, we estimate the degree of non-Markovianity using a classical kernel, i.e. the radial basis function (RBF). We follow the procedure reported in Ref.~\cite{Fanchini2021}, where the training is carried out with the expectation values ($\sigma_x,\sigma_y,\sigma_z$). Thus, instead of using quantum states to build up a kernel, we resort to use classical data, i.e. measurement outcomes. However, the process to obtain the states to be measured is the same we outlined in section~\ref{Sec_DQS}--- in Ref.\cite{Fanchini2021} the authors used a master equation approach instead of digital quantum simulation. 

In Table~\ref{Table1} we show the mean squared errors for each model for the amplitude damping (AD) and phase damping (PD) channels. We remark that the quantum versions, those where the kernel is calculate from the overlap between quantum states, deliver accurate predictions that are comparable with the classical models, albeit we found that SVM with a RBF kernel provides the best accuracy, as evidenced in terms of the mean squared error and the coefficient of determination $R^2$ (not shown here). This particular problem illustrates that extending the kernel to be quantum provides interesting insights and contributes to concatenate quantum blocks of operations. It not necessarily outperforms a fully classical training process but delivers useful outcomes.     

\begin{table}
\caption{\label{Table1}The Table shows the accuracy of the quantum and classical version of the studied machine learning models. The hyperparameters for AD (PD) are, QSVM: $ C=4\times10^{-1} (2\times10^{-1}),\epsilon=10^{-2}$; QKRR: $ \alpha=10^{-1} (2\times10^{-1})$; SVM: $ C=10^2,\epsilon=10^{-3}$; KRR: $\alpha=10^{-4} (10^{-5})$.}
\begin{ruledtabular}
\begin{tabular}{cccccc}
  & \mbox{QSVM}  & \mbox{QKRR} & \mbox{SVM} & \mbox{KRR}\\
\hline
AD & $6.0\times10^{-5}$   & $2.7\times10^{-5}$ & $2.6\times10^{-6}$ & $1.4\times10^{-5}$ \\
PD & $3.3\times10^{-4}$   & $1.6\times10^{-4}$ & $5.9\times10^{-5}$ & $1.8\times10^{-4}$
\end{tabular}
\end{ruledtabular}
\end{table}

\section{Conclusions}\label{Sec_conclusions}

In this paper we have thoroughly studied kernel-based quantum machine learning models to predict the degree of non-Markovianity using quantum data (quantum states). Each state is obtained through digital quantum simulation, where an ancilla qubit originates the non-Markovian behavior. We focus on two different decoherence channels, amplitude damping and phase damping. These quantum states are mapped to a Gram matrix by calculating its overlap. We investigate different kernel functions, say: linear, polynomial and exponential, and different kernel circuits to compute the overlap, say: inversion test, bell-basis algorithm, ancilla-based algorithm and the Swap test. We found that the inversion test with the exponential function delivers the best results. We draw our attention to two well-known kernel based machine learning models, Support Vector Machine (SVM) and Kernel Ridge (KRR). Because of their used with a precomputed quantum kernel we dubbed them as quantum SVM (QSVM) and quantum KRR (QKRR), respectively. By optimizing the learning process through cross-validation steps and grid search we found a good accuracy in our models. We found QSVM to be slightly better than QKRR, not only in the prediction's accuracy, but also in requiring less training samples.  

Finally, we compare our results with their classical counterpart, i.e. when using classical data (expectation values) to train the models. While there are not significant differences, we observe that SVM with an RBF kernel delivers the best performance. This means that in this particular case it is better to measure upon the system and then process the measurement outcomes with machine learning techniques.

\section{acknowledgments} D.T. acknowledges support from Universidad Mayor through the Doctoral fellowship. A.N. acknowledges financial support from Fondecyt Iniciaci\'on No. 11220266. 

\appendix

\section{Lagrangian calculations with SVM}\label{Appendix_svm}

We begin with the Lagrangian in Eq.~\eqref{lagrangian},
\begin{align}
L &=\frac{1}{2}\left\Vert \vec{w}\right\Vert^2 + C\sum_i \left( \xi_i + \xi_i^\ast \right) - \sum_i (\eta_i\xi_i +\eta_i^\ast\xi_i^\ast)\nonumber \\
& -\sum_i\alpha_i(\epsilon + \xi_i -y_i + \vec{w}\cdot\vec{x}_i +b)\nonumber \\
& -\sum_i\alpha_i^\ast (\epsilon + \xi_i^\ast + y_i -\vec{w}\cdot\vec{x}_i -b).
\end{align}

Taking the partial derivatives with respect to the primal variables ($b,w,\xi_i,\xi_i^\ast$) yields,

\begin{align}
&\partial_b L=\sum_i(\alpha_i^\ast-\alpha_i)=0, \\
&\partial_w L=w-\sum_i(\alpha_i-\alpha_i^\ast)x_i=0, \\
&\partial_\xi=C-\alpha_i-\eta_i=0,\label{partial_xi}\\
&\partial_{\xi^\ast}=C-\alpha_i^\ast-\eta_i^\ast=0.
\end{align}

First, from the KKT condition we obtain $\eta_i\xi_i=0$. By multiplying Eq.~\eqref{partial_xi} with $\xi_i$, we deduce the relation as,
\begin{equation}
(C-\alpha_i)\xi_i=0.
\end{equation}

This means that only samples with $\alpha_i=C$ lie outside the $\epsilon$-tube ($\xi_i\neq0$). We now consider the second constraint,
\begin{equation}
\alpha_i(\epsilon + \xi_i -y_i + \vec{w}\cdot\vec{x}_i +b)=0.
\end{equation}

Note that all samples inside the $\epsilon$-tube ($|f(\vec{x}_i)-y_i|<\epsilon$) have a vanishing Lagrange multiplier $\alpha_i$, which leads to the sparse representation of $f(\vec{x})$ in Eq.~\eqref{f_decision}. A similar procedure can be followed for $\xi_i^\ast,\eta_i^\ast,\alpha_i^\ast$, which allows to approach the value for $b$~\cite{Smola2004}.  

\section{Kernel functions performance}\label{Appendix_kernel_func}

\begin{figure}[ht]
	\centering
	\includegraphics[width=0.45 \textwidth]{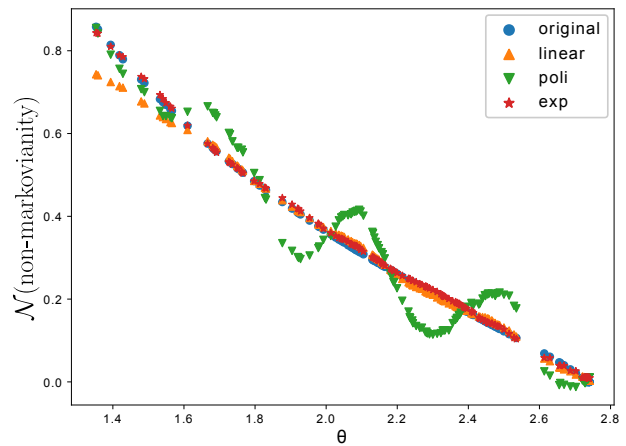}
	\caption{Exponential kernel function delivers the best prediction of non-Markovianity.}
	\label{fig:kernel_func}
\end{figure}

We now compare three different functions for the kernel $k(x_i,x_j)$, say: linear $\langle x_i,x_j\rangle$, polynomial $\left(\langle x_i,x_j\rangle +0.1\right)^3$, and exponential $\exp(-3\sqrt{1-\langle x_i,x_j\rangle})$. Figure~\ref{fig:kernel_func} shows that the exponential kernel function provides the best fitting. The polynomial function is only considered for completeness, since a more thorough exploration of the parameters may lead to a better fitting. 

\end{document}